\documentclass[10pt,twocolumn,letterpaper]{article}

\usepackage[pagenumbers]{ieee} % To force page numbers, e.g. for an arXiv version

\usepackage{algorithm}
\usepackage{algpseudocode}
\usepackage{mathtools}
\usepackage{multirow}
\usepackage{graphicx}

\usepackage{pdflscape}
\usepackage{array}

\usepackage[inline]{enumitem}   % package for within-paragraph lists

\usepackage{xcolor}             % to highlight text
\usepackage{soul}

\usepackage{pifont}
\definecolor{ieeeblue}{rgb}{0.21,0.49,0.74}
\usepackage[breaklinks,colorlinks,allcolors=ieeeblue]{hyperref} %Remove pagebackref

\title{UltraGauss: Ultrafast Gaussian Reconstruction of 3D Ultrasound Volumes}

\author{
Mark C. Eid$^{\, 1, 2, \, \raisebox{-0.25ex}{\href{mailto:markeid@robots.ox.ac.uk}{\includegraphics[width=0.7em]{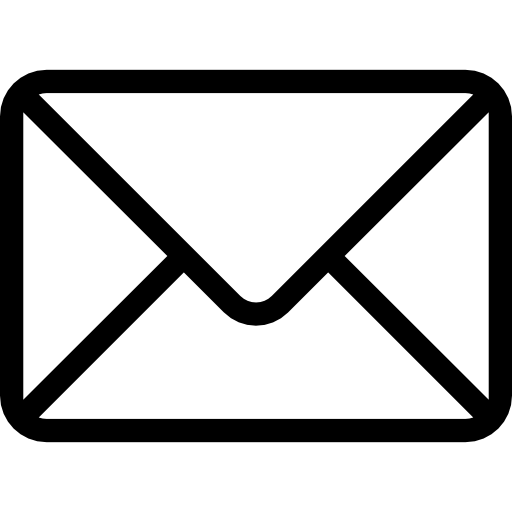}}}}$ \and Ana I.L. Namburete$^{2}$ \and João F. Henriques$^{1}$ \and  \\
$^{1}$Visual Geometry Group, University of Oxford \\
$^{2}$Oxford Machine Learning in NeuroImaging Lab, University of Oxford \\
$^{\raisebox{-0.25ex}{\includegraphics[width=0.7em]{mail.png}}}\,$\href{mailto:markeid@robots.ox.ac.uk}{markeid@robots.ox.ac.uk}
}

\begin{document}
\maketitle
\begin{abstract}
Ultrasound imaging is widely used due to its safety, affordability, and real-time capabilities, but its 2D interpretation is highly operator-dependent, leading to variability and increased cognitive demand. 
2D-to-3D reconstruction mitigates these challenges by providing standardized volumetric views, yet existing methods are often computationally expensive, memory-intensive, or incompatible with ultrasound physics.
We introduce UltraGauss: the first ultrasound-specific Gaussian Splatting framework, extending view synthesis techniques to ultrasound wave propagation. 
Unlike conventional perspective-based splatting, UltraGauss models probe-plane intersections in 3D, aligning with acoustic image formation. 
We derive an efficient rasterization boundary formulation for GPU parallelization and introduce a numerically stable covariance parametrization, improving computational efficiency and reconstruction accuracy.
On real clinical ultrasound data, UltraGauss achieves state-of-the-art reconstructions in 5 minutes, and reaching 0.99 SSIM within 20 minutes on a single GPU. 
A survey of expert clinicians confirms UltraGauss’ reconstructions are the most realistic among competing methods. 
Our CUDA implementation will be released upon publication.
\end{abstract}    
\section{Introduction}
\label{sec:intro}

\begin{figure}[t]
    \centering
    \includegraphics[width=1\linewidth]{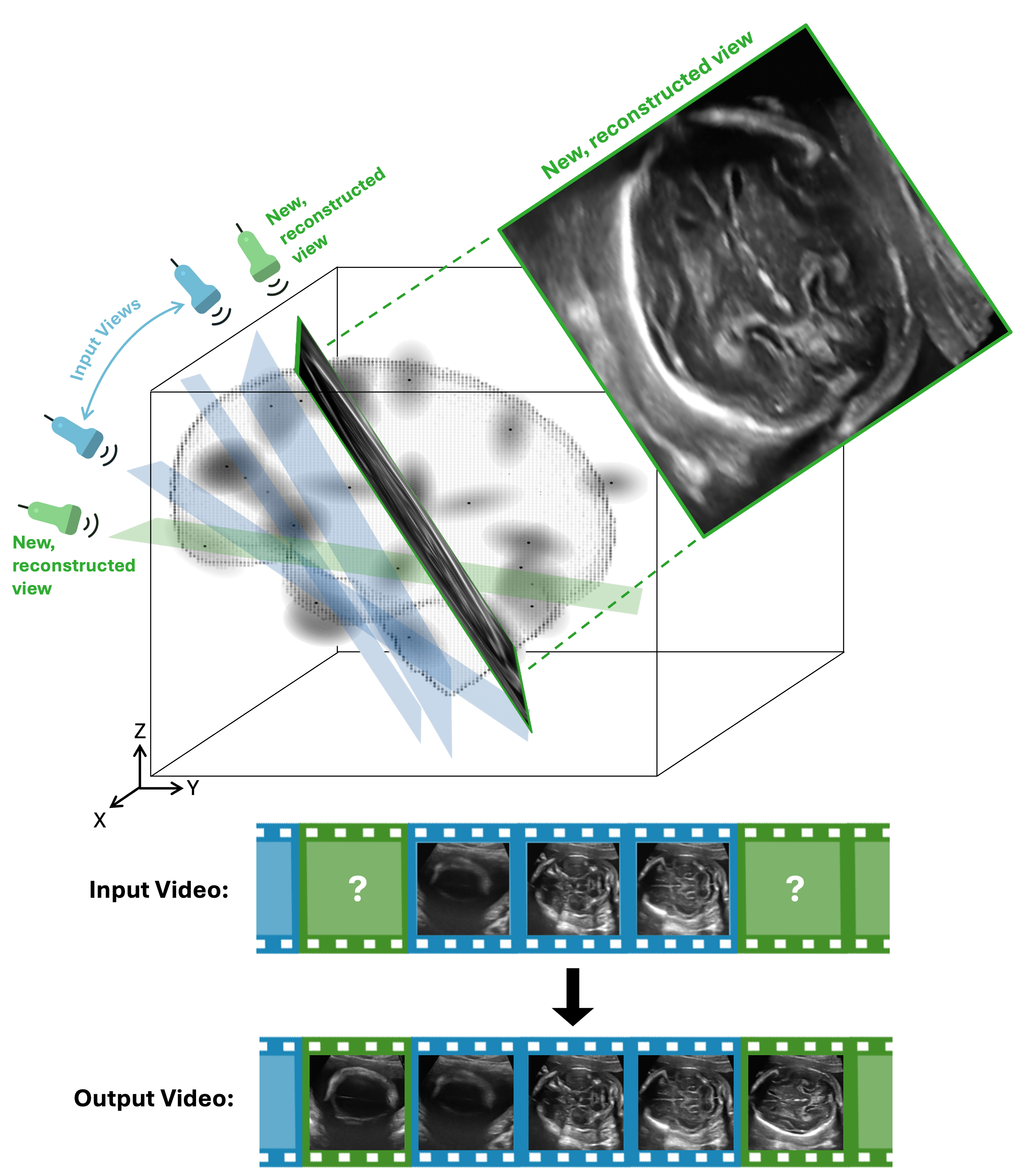}
    \caption{Input images of the fetal brain are acquired (blue Input Video frames) and their poses estimated (blue probes/planes). UltraGauss then uses these to form a total 3D reconstruction within 5 minutes. A few of the 3D gaussians are shown as grey ellipses. Cross-sectional views at previously unseen poses can then be sampled from the reconstructed 3D volume to form a complete cinesweep or be viewed individually. These can be seen in the green Output Video cinefilm frames.}
    \label{fig:us video splash}
\end{figure}

\begin{figure}[t]
    \centering
    \includegraphics[width=0.95\linewidth]{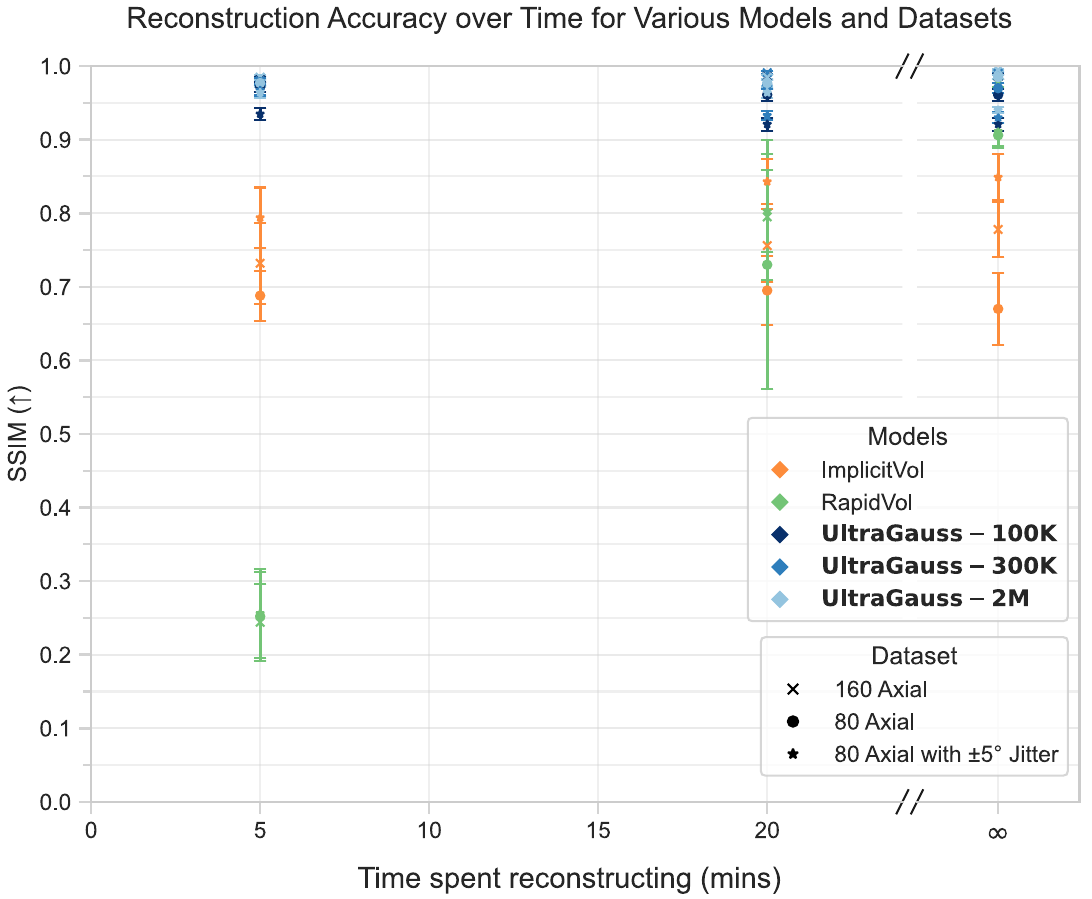}
    \caption{Reconstruction results for 5 different models (ours in \textbf{bold}) applied on 3 different training datasets, and shown at 3 time points over the duration of reconstruction. Higher SSIM is better. Errors bars show $\pm 1$ standard deviation amongst fetuses}
    \label{fig:main experiemnt results graph}
\end{figure}

Ultrasound (US) is one of the most widely used medical imaging modalities due to its real-time capabilities, portability, and safety (thanks to its non-ionizing nature).
However, clinical interpretation relies on sonographers mentally reconstructing 3D anatomies from 2D cross-sectional views -- a process that demands significant expertise \cite{benacerraf2002, nelson1998}.
This cognitive process, while effective for experienced sonographers, introduces observer-dependent variability and subjectivity, thus limiting reproducibility and standardization in clinical assessments. 3D US probes would be ideal, however they are around 10$\times$ more expensive than conventional 2D US probes, and require additional operator training, so their clinical use is extremely limited \cite{merz_2d_2005}. Automated 2D-to-3D ultrasound reconstruction is therefore essential to enhance spatial understanding, quantitative volumetric analysis, and improve diagnostic consistency.

Existing 2D-to-3D reconstruction methods suffer from computational inefficiency, memory constraints, or incompatibility with ultrasound physics.
Traditional techniques rely on segmentation, registration, and interpolation \cite{sarmah_survey_2023}, but struggle with irregular sampling and require extensive manual intervention.
More recent deep learning-based methods, such as ImplicitVol \cite{yeung_implicitvol_2021} and RapidVol \cite{eid_rapidvol_2024}, leverage \textit{implicit}, NeRF-like neural representations for higher fidelity reconstructions.
However, fully implicit methods are computationally expensive, while \textit{explicit} voxel-based methods suffer from memory overhead and resolution constraints.
Moreover, state-of-the-art methods often assume light-based ray accumulation, which is a fundamental mismatch with ultrasound physics, where wave propagation dictates image formation \cite{powles_physics_2018,aldrich_basic_2007}.

In this work we introduce \textbf{UltraGauss}: a novel volumetric ultrasound reconstruction framework designed to align with ultrasound wave physics rather than optical ray-based rendering.
Unlike conventional Gaussian Splatting (GS), which projects 3D Gaussians onto a 2D camera image plane for view-dependent rendering \cite{kerbl_3d_2023}, UltraGauss leverages anisotropic Gaussians to efficiently model ultrasound-specific wave propagation \cite{szabo2014}.
This fundamental distinction necessitates an in-plane approach, where GS is optimized to represent reconstructed wave reflections, as the ultrasound signal traverses and interacts with heterogeneous tissues.
This enables faster and more memory-efficient reconstruction while preserving high-fidelity volumetric representations.

We demonstrate the versatility of UltraGauss in multiple scenarios, achieving high-fidelity 3D reconstructions from axial slices, and delivering consistent structural accuracy across orthogonal axial, coronal, and sagittal views. 
Additionally, we benchmark it against prior neural reconstruction methods for ultrasound, validate its robustness using noisy and partially sampled slices, and assess its real-world applicability by reconstructing volumes from real 2D cine-sweep videos (\cref{fig:us video splash}). 
Our method achieves an SSIM of 0.91 on unseen video frames, with expert clinicians (mean = 18 years of medical experience)
confirming its reconstructions are visually indistinguishable from the ground truth (that is, native 3D acquisitions).
UltraGauss achieves real-time rendering speeds, and can obtain a SOTA reconstruction in 5 minutes on a single GPU, making it highly suitable for clinical and research applications.

In summary, our technical contributions are:
\begin {itemize} %[label=\itshape\alph*\upshape)]
\item An ultrasound-specific 2D-to-3D reconstruction framework, adapting and optimising anisotropic gaussians for wave-based ultrasound physics rather than light-based projection techniques.
\item An efficient triangular covariance parametrization, which is more numerically stable than that used in prior work, and which facilitates efficient inversion and factorization, operations that are frequently used during reconstruction.
\item A load-balancing scheme to optimally distribute computation across parallel threads, while simultaneously rejecting gaussians that do not contribute to rendering.
\item A comprehensive validation of clinical ultrasound datasets, demonstrating significant speed improvements over state-of-the-art methods while maintaining high-fidelity volumetric reconstructions.
\item A qualitative survey with expert clinicians, which shows that our reconstructions are deemed more realistic in a clinically-relevant setting.
\end{itemize}

\section{Related Work}

\paragraph{2D-to-3D reconstruction in medical imaging:}
2D-to-3D reconstruction is a fundamental problem in medical imaging, enabling volumetric representation from cross-sectional slices acquired by X-ray, ultrasound, CT, or conventional cameras.
Traditional approaches include segmentation-based methods, where 2D slices are first segmented (e.g., via pixel thresholding), spatially registered using B-splines, and then interpolated into a 3D volume \cite{sarmah_survey_2023}.
However, these methods require extensive manual intervention, struggle with irregular sampling, and scale poorly to high-resolution datasets.

More recently, deep learning approaches have transformed 3D reconstruction by leveraging CNN-based segmentation and implicit neural representations \cite{cao_image_2020}.
Specifically, volumetric representations have evolved from regular voxel grids and octrees to more memory-efficient point clouds and implicit neural fields \cite{yuniarti_review_2019,dalal_gaussian_2024}.
Classical methods such as Structure-from-Motion (e.g., COLMAP \cite{schoenberger2016sfm}) reconstruct 3D surfaces by triangulating depth from multiple geometrically distinct views.
Although ultrasound can be collected from different angles, it is often impractical to acquire the broad range of precisely registered viewpoints required for accurate SfM-based 3D reconstruction.
Furthermore, SfM assumes photometric consistency, which does not hold for ultrasound due to wave-based interactions, such as refractions and signal attenuation.

\noindent \textbf{NeRFs and Gaussian Splatting:}
NeRFs \cite{mildenhall_nerf_2020} introduced a paradigm shift in 3D reconstruction by encoding volumetric scenes within a neural network.
In brief, NeRFs represent 3D space as a learnable function $F_\Theta (x,y,z,\theta,\phi)$ that maps spatial coordinates and viewing directions to color and density values.
While NeRFs enable high-fidelity novel view synthesis, they rely on ray marching and volume accumulation, making them computationally intensive and unsuited for ultrasound \cite{aldrich_basic_2007}. 

A breakthrough came with Gaussian Splatting \citep{kerbl_3d_2023}, which replaces implicit NeRF-based volume rendering with explicit 3D Gaussian representations.
Instead of marching rays through a neural field, GS represents a scene as a collection of 3D Gaussians, which are rasterized (``splatted'') onto the 2D image plane, accumulating color and opacity in a differentiable manner.
This approach has led to medical imaging applications in endoscopy, where it enables dynamic 3D scene reconstruction from monocular endoscopic videos using depth-guided optimization \cite{linguraru_free-surgs_2024}, and CT-based reconstruction, where it facilitates sparse-view tomographic reconstruction by improving volume estimation through learned 3D Gaussian representations \cite{zha_r2-gaussian_2024}.
However, these methods remain projection-based, assuming light transport.

\noindent \textbf{Ultrasound-specific challenges:}
Unlike optical imaging, ultrasound does not accumulate intensity along a ray but instead propagates as (mechanical) sound waves that interact with biological tissues \cite{powles_physics_2018,aldrich_basic_2007}.
This makes NeRF-style ray tracing and projection-based GS fundamentally unsuitable for ultrasound reconstruction.
Traditional explicit voxel grids are memory-intensive and resolution-limited \cite{solberg_freehand_2007}, while implicit representations such as ImplicitVol \cite{yeung_implicitvol_2021} and RapidVol \cite{eid_rapidvol_2024} improve scalability but remain computationally expensive.

To the best of our knowledge, no ultrasound-adapted GS method exists for 2D-to-3D reconstruction.
Our proposed approach, UltraGauss, extends GS to volumetric ultrasound by replacing projection-based rendering with a model that aligns with ultrasound wave propagation.
By leveraging anisotropic Gaussians, UltraGauss enables faster, memory-efficient, and clinically relevant 3D ultrasound reconstruction, without requiring external probe tracking systems.

\section{Background}
\label{sec:background}
\begin{figure}
    \centering
    \includegraphics[width=0.7\linewidth]{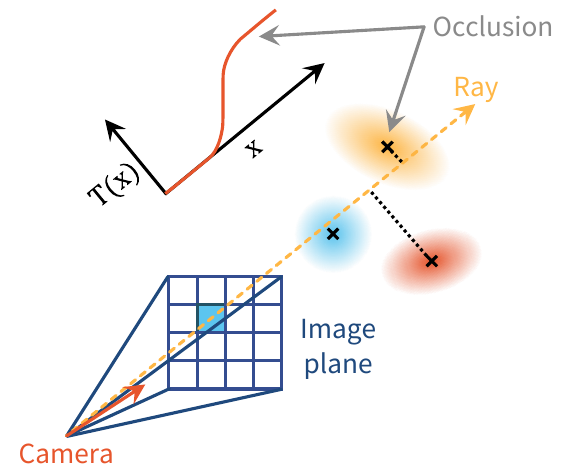}
    \caption{Illustration of the occlusion mechanism in gaussian splatting, for images in the visual spectrum. The rendering equation (\cref{eq:rgb-render}) accumulates transmittance $T$ over a pixel's ray, and this is used to model occluding gaussians (in this example, the blue gaussian occludes the yellow one).}
    \label{fig:rgb-gauss}
\end{figure}

For cameras in the visible spectrum, one can render the color $c_{\mathrm{RGB}}$ for a given pixel by evaluating a volumetric model along the corresponding ray, sampling the model's opacities $\hat{\alpha}(x)$ and colors $\hat{c}(x)$ at $m$ ordered sampled points $x_{j}$ on the ray:
\begin{equation}
c_{\mathrm{RGB}}=\sum_{j=1}^{m}T_{j}\hat{\alpha}\left(x_{j}\right)\hat{c}\left(x_{j}\right),\quad T_{j}=\prod_{k=1}^{j-1}\left(1-\hat{\alpha}\left(x_{k}\right)\right),\label{eq:rgb-render}
\end{equation}
where $T_{j}$ denotes the accumulated transmittance of the material. The main purpose of the accumulated transmittance is to model occlusions (see \cref{fig:rgb-gauss}).
In the case of gaussian splatting \cite{kerbl_3d_2023}, the volumetric model is a weighted sum of $n$ gaussian functions with means $\mu_{i}$, covariances $\Sigma_{i}$, colors $c_{i}$, and coefficients (maximum opacities) $\alpha_{i}$. In contrast to NeRFs \cite{mildenhall_nerf_2020}, the sampled points $x_{j}$ can now be reduced to the projections of close gaussians on to the ray, which can be made more efficient. The opacity of the $i$th gaussian at a 2D (image-space) point $x$ is then calculated as
\begin{equation}
\hat{\alpha}_{i}\left(x\right)=\alpha_{i}\exp\left(-\frac{1}{2}\right.\underbrace{\left(x-\mu_{i}^{\mathrm{2D}}\right)^{T}\left(\Sigma_{i}^{\mathrm{2D}}\right)^{-1}\left(x-\mu_{i}^{\mathrm{2D}}\right)}_{\textrm{2D squared Mahalanobis distance}}\left.\vphantom{\frac{1}{2}}\right),\label{eq:mahalanobis-2d}
\end{equation}
with the mean $\mu_{i}^{\mathrm{2D}}$ and covariance $\Sigma_{i}^{\mathrm{2D}}$ projected to the 2D image. To achieve this, the 3D parameters $\left(\mu_{i},\Sigma_{i}\right)$ are translated and rotated to the camera's reference frame by a view transformation (extrinsic camera matrix) $W$, and projected to 2D with the affine approximation of a projective transformation (using intrinsic camera matrix $K$) \cite{zwicker_ewa_2001}. We can express this formally with operators to convert to and from homogenous coordinates, $\hbar(u)=\left[u_{1},\ldots,u_{D},1\right]^{T}$ and $\hbar^{-1}(u)=\left[u_{1}/u_{D},\ldots,u_{D-1}/u_{D}\right]$ respectively:
\begin{equation}
\mu_{i}^{\mathrm{2D}}=\mathrm{proj}\left(\mu_{i}\right)=\hbar^{-1}\left(KW\hbar\left(\mu_{i}\right)\right)
\end{equation}
\begin{equation}
\Sigma_{i}^{\mathrm{2D}}=J_{i}W\Sigma_{i}W^{T}J_{i}^{T},\quad J_{i}=\frac{\partial\mathrm{proj}(\mu_{i})}{\partial\mu_{i}}
\end{equation}
We now opt to include an additional uniform component (background) with color $c_{\mathrm{BG}}$ and coefficient $\alpha_{\mathrm{BG}}$, which improves numerical stability by avoiding a division by zero. We can then combine the opacities and colors of all components to obtain the rendered opacity $\hat{\alpha}\left(x\right)$ and color $\hat{c}\left(x\right)$ for \cref{eq:rgb-render}:
\begin{alignat}{1}
\hat{\alpha}\left(x\right) & =\sum_{i}^{n}\hat{\alpha}_{i}\left(x\right)+\alpha_{\mathrm{BG}}\label{eq:alpha-combine}\\
\hat{c}\left(x\right) & =\frac{1}{\hat{\alpha}\left(x\right)}\left(\sum_{i}^{n}\hat{\alpha}_{i}\left(x\right)c_{i}+\alpha_{\mathrm{BG}}c_{\mathrm{BG}}\right)\label{eq:colour-combine}
\end{alignat}
In this summary we leave out spherical harmonics, which support directionally-dependent colors \cite{ramamoorthi_modeling_2006, yu_plenoxels_2021}.

\subsection{Optimization\label{subsec:background-optimization}}
While \crefrange{eq:rgb-render}{eq:colour-combine} can be used to render a gaussian model, in practice several optimizations are necessary to avoid the costly nested iterations that they imply. Namely, the gaussians are assigned to tiles and sorted by depth, and distributed across GPU threads for rasterizing into each pixel \cite{kerbl_3d_2023}. Depth sorting and cut-off distances for the gaussians in the image plane make the rendering process only approximate \cite{Huang2DGS2024}. Finally, several heuristics are necessary to remove or resample gaussians in areas that are too sparse or too dense \cite{rota_bulo_revising_2025, yu_mip-splatting_2023}.

\begin{figure}
    \centering
    \includegraphics[width=0.7\linewidth]{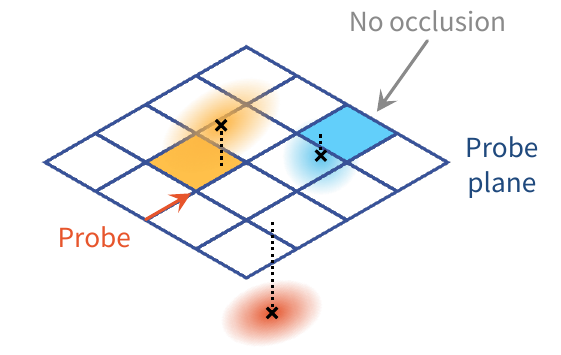}
    \caption{Illustration of the proposed gaussian rendering mechanism for ultrasound (\cref{sec:image-formation}). Instead of raycasting, the image formation model is based on the intersection of the gaussians with the probe plane (which maps 1:1 to the captured ultrasound pixel grid). This means that occlusions do not play as large a role as for the visual spectrum (\cref{fig:intersection}), and so are not modeled.}
    \label{fig:ultrasound-gauss}
\end{figure}

\section{Method}

\subsection{Image formation model -- ultrasound vs. visual spectrum}\label{sec:image-formation}

While fast gaussian splatting is well-developed for RGB image cameras, there are important differences in the image formation model of ultrasound data that require rethinking many of the design choices. The main reason is that ultrasound probes use reflected ultrasound waves to measure the response of materials at a dense range of depths from the probe (\cref{fig:ultrasound-gauss}). Image cameras, in contrast, measure reflected light that typically bounces off solid objects, and thus are more affected by occlusions and most often measure surfaces (\cref{fig:rgb-gauss}).

This means that the accumulated transmittance $T_{j}$ in \cref{eq:rgb-render} is incorrect, since it will treat opaque volumes as occlusions (see $T(x)$ subplot in \cref{fig:rgb-gauss}). Instead, the main rendering mechanism for ultrasound images must be to detect \emph{intersections} with the probe plane (\cref{fig:ultrasound-gauss}), not projections.

Formally, we can have a simpler rendering equation for ultrasound images, by reusing the definition of $\hat{c}\left(x\right)$ from \crefrange{eq:alpha-combine}{eq:colour-combine} to combine the gaussian and background components at a 2D point (pixel) $x$:
\begin{equation}
c_{\mathrm{Ultrasound}}(x)=\hat{c}\left(x\right).\label{eq:us-render}
\end{equation}
A large conceptual difference from rendering the visual spectrum is that, instead of projecting the gaussian parameters to 2D, we need them to \emph{intersect} the probe plane in 3D. The opacity $\hat{\alpha}_{i}\left(x\right)$ of the $i$th gaussian at a 2D point $x$ (necessary for \crefrange{eq:alpha-combine}{eq:colour-combine}) is then evaluated as a Mahalanobis distance in 3D space, by \emph{lifting the 2D image point to 3D} in the coordinate-frame of the probe:
\begin{equation}
\hat{\alpha}_{i}\left(x\right)=\alpha_{i}\exp\left(-\frac{1}{2}\right.\underbrace{\left(x_{|0}-\mu_{i}^{\mathrm{3D}}\right)^{T}\left(\Sigma_{i}^{\mathrm{3D}}\right)^{-1}\left(x_{|0}-\mu_{i}^{\mathrm{3D}}\right)}_{\textrm{3D squared Mahalanobis distance}}\left.\vphantom{\frac{1}{2}}\right),\label{eq:mahalanobis-3d}
\end{equation}
where $x_{|0}=\left[x_{1},x_{2},0\right]^T$ and the gaussian's parameters are moved to the probe's coordinate-frame using its inverse transform matrix $W$:
\begin{equation}
\mu_{i}^{\mathrm{3D}}=\hbar^{-1}\left(W\hbar\left(\mu_{i}\right)\right),\quad\Sigma_{i}^{\mathrm{3D}}=W\Sigma_{i}W^{T}.
\end{equation}
Contrast \cref{eq:mahalanobis-2d} to \cref{eq:mahalanobis-3d}: despite the similarities, the former (for RGB images) evaluates a 2D gaussian after projecting it to image-space, while the later evaluates a 3D gaussian by doing the reverse operation (for ultrasound images). Using this model, we can now design a fast splatting strategy to avoid the computational expense of simply evaluating \crefrange{eq:alpha-combine}{eq:colour-combine} and \cref{eq:mahalanobis-3d} (which would result in nested iterations over all pixels and all gaussians).

\subsection{Triangular covariance parameterization for efficient inversion and gaussian sampling}

One challenge in optimizing representations with covariance matrices $\Sigma$ (omitting the subscript $i$ for conciseness) is that they must remain positive-definite (PD, all eigenvalues strictly positive), while gradient-based optimization methods typically only support unconstrained optimization. Kerbl et al. \cite{kerbl_3d_2023} achieved this by reparameterizing the covariances as a product of a scaling vector $s$ and a quaternion-derived rotation matrix $R$, as $\Sigma=R\,\mathrm{diag}\left(s^{2}\right)R^{T}$. However, empirically we found that, for our setting, the normalization of the quaternion (which must be unit-norm to represent a 3D rotation) resulted in numerical instabilities. Instead, we can ensure a matrix is PD by parameterizing it as a product of a matrix $M$ with itself (which is positive-semidefinite) and adding a small multiple of the identity $I$ to ensure positive eigenvalues (with $\beta>0$):
\begin{equation}
\Sigma'^{-1}=MM^{T}+\beta I,\quad M=\begin{bmatrix}M_{11} & M_{12} & M_{13}\\
M_{12} & M_{22} & M_{23}\\
M_{13} & M_{23} & M_{33}
\end{bmatrix}
\end{equation}
Note that $M$ itself is symmetric, to achieve the minimal number of degrees of freedom of a 3D covariance (6). No normalization is needed.
A remaining challenge is that, in addition to requiring the inverse covariance $\Sigma^{-1}$ frequently to render pixels (\cref{eq:mahalanobis-3d}), we also occasionally (e.g. every 100 iterations) must perform heuristic resampling of some gaussians. This operation requires \emph{inverting} $\Sigma^{-1}$\emph{ explicitly} to obtain the original covariance $\Sigma$, as well as factorizing it to draw a sample from the multivariate gaussian (\cref{subsec:background-optimization}). Both operations can be numerically unstable for ill-conditioned $\Sigma^{-1}$. Therefore, we propose instead a more efficient parameterization, as a product of a lower-triangular matrix $L$:
\begin{equation}
\Sigma^{-1}=LL^{T},\quad L=\begin{bmatrix}L_{11}^{2}+\beta & 0 & 0\\
L_{12} & L_{22}^{2}+\beta & 0\\
L_{13} & L_{12} & L_{33}^{2}+\beta
\end{bmatrix}.\label{eq:tri-param}
\end{equation}
\Cref{eq:tri-param} guarantees that $\Sigma^{-1}$ is PD, since a triangular matrix's eigenvalues are its diagonal elements, and it can be seen that these are strictly positive (with $\beta>0$). Moreover, a lower-triangular matrix is extremely efficient to invert via \emph{forward substitution}, which is implemented in most numerical packages. This allows easily computing $\Sigma=\left(L^{-1}\right)^{T}L^{-1}$. Finally, another advantage is that this factorization allows sampling from a gaussian (for the resampling heuristic, \cref{subsec:background-optimization}), by simply projecting a standard normal sample $z$ with the (efficiently) inverted $L$:
\[
y=\mu+L^{-T}z,\quad z\sim\mathcal{N}(0,1).
\]

Even though \cref{eq:tri-param} supports efficient calculation of the opacity $\hat{\alpha}_{i}\left(x\right)$ (\cref{eq:mahalanobis-3d}) and resampling heuristics, for most pixel-gaussian pairs the opacity will be close to 0 and can be ignored. The next section will focus on this aspect.

\begin{figure}
    \centering
    \includegraphics[width=0.8\linewidth]{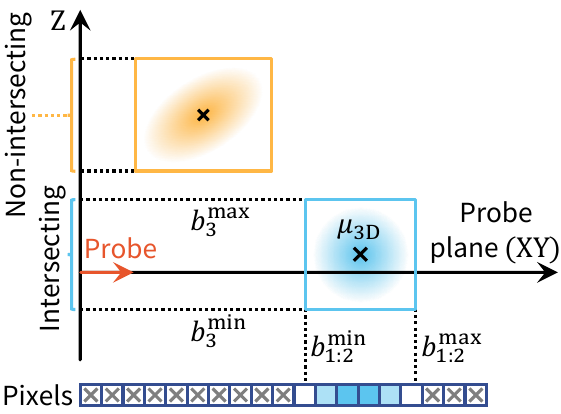}
    \caption{Side view of the gaussian bounding box intersection and projection (\cref{sec:boundaries}). Boxes that do not intersect with the probe plane are rejected early (\cref{sec:load-balance}). Those that intersect it are rasterized only within the intersecting 2D bounding box. In this example, pixels that are not iterated over are marked with crosses.}
    \label{fig:intersection}
\end{figure}

\subsection{Rasterization boundaries}\label{sec:boundaries}
To avoid evaluating \cref{eq:mahalanobis-3d} when it is known to yield opacities close to 0, we compute it only inside a bounding box for each gaussian. The bounding box is defined around the ellipsoid that encompasses $p\%$ of the gaussian's probability density. This ellipsoid can be expressed as a function of the 3D squared Mahalanobis distance from \cref{eq:mahalanobis-3d}, and the chi-squared distribution value for 3 degrees of freedom (omitting the subscript $i$ for clarity):
\begin{equation}
\left(x_{|0}-\mu^{\mathrm{3D}}\right)^{T}\left(\Sigma^{\mathrm{3D}}\right)^{-1}\left(x_{|0}-\mu^{\mathrm{3D}}\right)\leq\chi_{3,1-p}^{2},\label{eq:ellipsoid}
\end{equation}
which evaluates to $\chi_{3,1-p}^{2}=7.815$ for $p=95\%$ (note that it is a bound on a squared distance). Now to calculate the bounding box of this ellipsoid, its minimum and maximum coordinates are the 3D vectors $b^{\min}$ and $b^{\max}$:
\begin{equation}
b^{\min/\max}=\mu^{\mathrm{3D}}\pm\sqrt{\lambda v},\quad\lambda=\frac{\chi_{3,1-p}^{2}}{\det\left(\Sigma^{\mathrm{3D}}\right)},\label{eq:boundary}
\end{equation}
\[
v=\left[\begin{array}{c}
\Sigma_{22}^{\mathrm{3D}}\Sigma_{33}^{\mathrm{3D}}-\left(\Sigma_{23}^{\mathrm{3D}}\right){}^{2}\\
\Sigma_{11}^{\mathrm{3D}}\Sigma_{33}^{\mathrm{3D}}-\left(\Sigma_{13}^{\mathrm{3D}}\right){}^{2}\\
\Sigma_{11}^{\mathrm{3D}}\Sigma_{22}^{\mathrm{3D}}-\left(\Sigma_{12}^{\mathrm{3D}}\right){}^{2}
\end{array}\right],
\]
using $\Sigma_{jk}^{\mathrm{3D}}$ to index the $(j,k)$th element of a matrix. \Cref{eq:boundary} can be obtained by writing \cref{eq:ellipsoid} as a function of each element of $\delta=x_{|0}-\mu^{\mathrm{3D}}$.
\Cref{fig:intersection} illustrates these bounds.

\subsection{Load balancing across rendering threads}\label{sec:load-balance}

Note that the bounding box (\cref{eq:boundary}) can be partitioned into 2 components:
\begin{enumerate}
\item The 2D bounding box in the probe plane (first two elements of $b^{\min/\max}$, i.e. $b_{1:2}^{\min/\max}$).
\item The 1D segment orthogonal to the probe plane (the third element, $b_{3}^{\min/\max}$).
\end{enumerate}
This suggests a two-phase process for efficient rendering:
\begin{enumerate}
\item Reject any gaussians whose cut-off boundaries do not intersect with the probe plane: $b_{3}^{\min}>0$ or $b_{3}^{\max}<0$.
\item For each accepted gaussian, only iterate over 2D pixels $x$ inside the bounding box of the plane: $b_{1}^{\min}\leq x_{1}\leq b_{1}^{\max}$ and $b_{2}^{\min}\leq x_{2}\leq b_{2}^{\max}$.
\end{enumerate}
This process is visualized in \cref{fig:intersection}.

This provides a natural mechanism for load balancing across parallel threads of computation. Phase 1 requires iterating through all the gaussians (but not the pixels), marking them as accepted or rejected based on the perpendicular projection, which can be equally partitioned between all the threads. A standard operation of buffer compaction \cite{Thrust} can then reduce this list of gaussians to only the accepted ones.

Phase 2 then requires iterating through the compacted list of accepted gaussians, and rasterizing each one only onto the corresponding bounding box of the image buffer, by atomically adding to pixel accumulators for color $\hat{c}\left(x\right)$ and opacity $\hat{\alpha}\left(x\right)$ (implementing \crefrange{eq:alpha-combine}{eq:colour-combine}). The compacted list in Phase 2 can be equally partitioned among the threads, to ensure optimal throughput.

\subsection{Optimization}

Like gaussian splatting for light-based images, we use back-propagation and stochastic gradient descent to optimize the gaussian parameters ($\mu_{i}$, $\Sigma_{i}$, $c_{i}$, $\alpha_{i}$) \cite{kerbl_3d_2023}.
Unlike this prior work, we do not have access to a COLMAP-derived point cloud for initialization, which would only cover surfaces, not the interior of a full volume. Instead, we initialize the gaussians' means $\mu_{i}$ uniformly at random within the imaged volume boundaries, with $c_{i} = 0.5$ and $\alpha_{i} = 0.731$. $L{ij}$ in \cref{eq:tri-param} are uniformly randomly initialized between $[4,5)$, corresponding to a variance between 0.0017 and 0.0043.
We follow the original paper in all other training details, incorporating the same heuristics for sparsification and densification after some slight modification due to the lack of covariance ``splatting''/projection in UltraGauss. We avoid repeating the exact methods here for brevity, and refer interested readers to \cite{kerbl_3d_2023, ye_gsplat_2024} for more details. Noteworthy however is the starting number of gaussians. Due to the efficiency of our method and CUDA kernels, we are able to have up to 2 million gaussians (possibly even more) with still very respectable reconstruction times. As is to be expected, less gaussians leads to a slightly lower final accuracy, but results in a speed up. We thus try 3 different variants of our model in \cref{sec: Experiments}, having either \{$100{\times}10^3$, $200{\times}10^3$, $2{\times}10^6$\} gaussians. We optimize (train) all reconstruction models on a single NVIDIA RTX A4000 GPU, with Adam optimizers and learning rates (lr) of 0.05 for all gaussian parameters, except for the means $\mu_{i}$ which have an exponential decay scheduler (as in \cite{yu_plenoxels_2021, kerbl_3d_2023}) and a starting lr of 0.00016. Like Kerbl et al. \cite{kerbl_3d_2023} we use a sigmoid activation function to constrain $\alpha_{i}$ to the range $[0,1)$.
We implemented our method in PyTorch, with custom CUDA kernels for rasterization and its gradients. We will make our code available as open-source upon publication.

\section{Experiments}\label{sec: Experiments}

\subsection{Datasets} \label{subsec: Datasets}
Two clinical datasets were curated to validate UltraGauss in different ultrasound acquisition settings. \textbf{Dataset A} consists of volumetric ultrasound scans, allowing assessment of reconstruction fidelity across multiple orthogonal views. \textbf{Dataset B} comprises freehand ultrasound video sequences, evaluating reconstruction quality in the absence of full 3D coverage, mimicking real-world fetal monitoring scenarios.

\textbf{Dataset A: 3D Ultrasound Volumes}
This dataset includes twelve 3D fetal brain ultrasound volumes ($160 \times 160 \times 160$ voxels, $0.6 \times 0.6 \times 0.6$ $\text{mm}^3$ resolution), obtained from the INTERGROWTH-21$^\textrm{st}$ study. 
Acquisitions were performed between 14 and 26 gestational weeks: spanning a critical period of brain maturation \cite{namburete-brainage-2015,namburete_normative_2023}, and the standard time for fetal anomaly screening \cite{salomon_isuog_2022}. 
During this stage, rapid developmental changes result in varying structural appearance and composition across images \cite{namburete_normative_2023}. 
The scans were collected using a Philips HD9 curvilinear probe (2.5 MHz wave frequency) by multiple sonographers, which introduces variability in probe positioning and image appearance. 
Image dimensions varied across acquisitions, ranging from $117 \times 126 \times 126$ to $271 \times 343 \times 337$ voxels, with resolutions between $0.27 \times 0.69 \times 1.08$ mm and $0.46 \times 0.99 \times 1.83$ mm per voxel. 
To ensure consistency, all volumes were resampled to an isotropic resolution of $0.6 \times 0.6 \times 0.6$ $\text{mm}^3$ and resized to $160 \times 160 \times 160$ voxels.

\textbf{Dataset B: 2D Freehand Video Sequences}
Three freehand 2D ultrasound videos of fetal brain acquisitions were collected at 19 and 20 weeks’ gestational age at Leiden University Medical Center using a GE Voluson E10 ultrasound scanner. Each video consists of $\sim$100 frames, with each frame cropped and resized to $160 \times 160$ pixels, and resampled to a resolution of $0.6 \times 0.6$ $\text{mm}^2$.

\subsection{Evaluation of Reconstruction Quality and Speed} \label{subsec: Exp1}
To investigate the maximum potential of UltraGauss, we first evaluate it in ``ideal'' conditions, namely when given 160 linearly spaced axial slices sampled from a 3D Volume in Dataset A. This means that theoretically the entire $160 \times 160 \times 160$ voxel volume has been seen, so in theory perfect accuracy should be possible. When presented with a 3D US volume, whether acquired with a 3D US probe or reconstructed from a 2D probe, the standard way clinicians view it is by looking at a series of linearly spaced axial, coronal and sagittal cross-sections. We therefore evaluate our reconstruction in a similar manner, and sample 160 linearly spaced axial, coronal and sagittal views. As we have a native 3D US scan from Dataset A, we can also sample from that, giving us a ground truth to compare against. We then quantitively evaluate the quality of the reconstructed images using SSIM ($\uparrow$) \cite{wang_image_2004}, PSNR ($\uparrow$) and LPIPS ($\downarrow$) \cite{zhang_unreasonable_2018}. As the end-goal is for near real-time reconstruction, training time $t$ must be taken into account, and so we evaluate the quality of the reconstructions at $t=5$ and $t=20$ minutes as well as once convergence has reached (denoted with $t=\infty$).
We repeat the above when only 50\% of the volume is given, by firstly reconstructing on only 80 linearly spaced axial slices, and then with a random amount of rotation about the x axis and y axis ($\theta \sim U(-5^{\circ},+5^{\circ})$) added, mimicking the unsteady hand of the sonographer and the movement of the fetus as an ultrasound sweep is acquired in practice. 
We also benchmark all this against two SOTA \underline{ultrasound} reconstruction models: RapidVol \cite{eid_rapidvol_2024}, a hybrid implicit-explicit model, and ImplicitVol \cite{yeung_implicitvol_2021}, a fully implicit NeRF-like model.

\subsection{Clinicians’ Survey of Reconstruction Quality} \label{subsec: Exp2}

To evaluate the clinical realism and fidelity of UltraGauss' reconstructions, we conducted a survey with expert sonographers, comparing natively acquired 3D ultrasound images with reconstructions from RapidVol, ImplicitVol, and UltraGauss. This study focused on 3 axial planes (mid-axial, transthalamic, transventricular) and one mid-coronal plane, which are routinely assessed during the mid-gestation anomaly scan (20 weeks, ISUOG guidelines \cite{salomon_isuog_2022}). 
Survey data was de-identified, for patient confidentiality.

\noindent \textbf{Study design:}
Participants were presented with randomized image pairs and asked to select the most realistic scan, or indicate ``no preference'' if indistinguishable. 
To capture variations in expertise, respondents provided details on speciality training, years in practice, and confidence in fetal brain assessment.

The survey evaluated five different reconstruction models (ImplicitVol, RapidVol, UltraGauss-100K, UltraGauss-300K, UltraGauss-2M), on four fetal brain scans, and at three training time points ($t=\{5,20,\infty\}$). 
Each participant reviewed 10 randomized image pairs, with both question and image order shuffled to mitigate bias.

Additionally, we assessed how UltraGauss' reconstructions compare to real ultrasound scans over training time. A separate survey presented participants with nine versions of a mid-coronal image generated at $t=\{0.5,1,2,3,4,5,10,15,20\}$ minutes, alongside two identical ground-truth images. 
The input set consisted of 80 axial slices, covering only 50\% of the full 3D volume and completely orthogonal to the generated image. 
Participants were asked to classify each image as either a real ultrasound scan or an AI reconstruction. 
The randomized inclusion of two identical ground-truth images served as a control to assess the inherent variability in sonographer judgment.

\subsection{UltraGauss in an end-to-end clinical pipeline} \label{subsec: Exp3}
A key application of UltraGauss is enabling clinicians to capture a freehand video (cinesweep) with a standard sensorless 2D probe, and then reconstruct a full 3D volume. This enables retrospective visualization of poses missed or not explicitly captured during scanning. The acquired frames would first have their poses predicted using an ultrasound pose-estimation model (\eg \cite{linguraru_geometric_2024,di_vece_ultrasound_2024, yeung_adaptive_2022}), and then both the frames and predicted poses are fed into a suitable 3D reconstruction model.
We attempt this end-to-end pipeline, by using QAERTS \cite{linguraru_geometric_2024} for pose estimation, and then UltraGauss for reconstruction. We also benchmark against RapidVol. Since ground-truth 3D scans of the fetus in the cinesweeps are unavailable, direct quantitative evaluation is not possible. Instead, we adopt a cross-validation strategy, randomly partitioning the video into training (80\%) and testing (20\%) frames. After reconstruction, we retrieve frames at the predicted poses of the held-out test slices and compare them to the actual frames. Any observed error reflects both pose estimation and reconstruction inaccuracies.
We test this pipeline on all three fetal cinesweeps in Dataset B.

\begin{figure}
    \centering
    \includegraphics[width=0.99\linewidth]{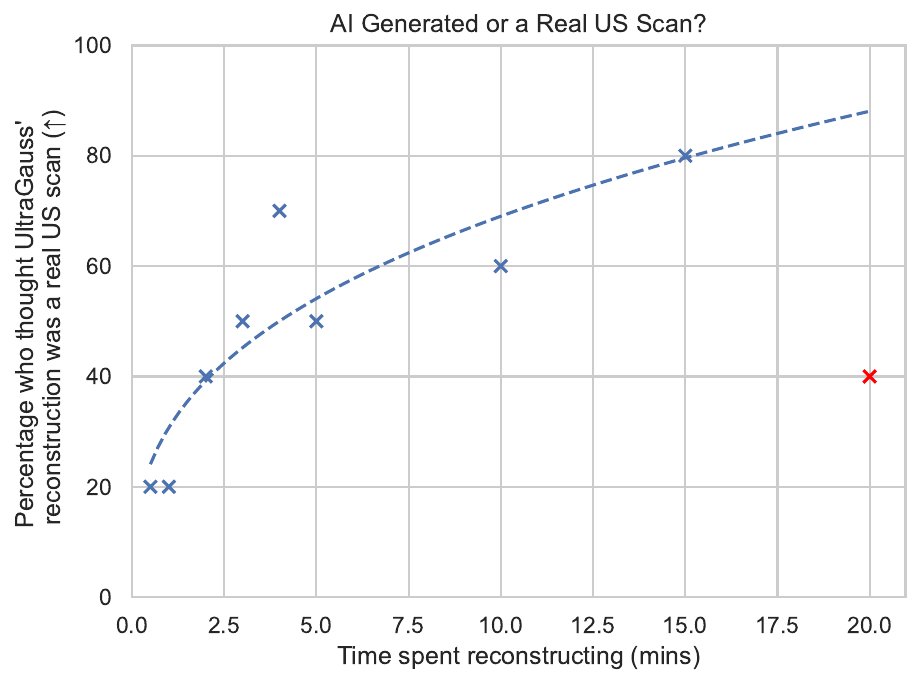}
    \caption{Results to the 2nd part of the Clinicians' Survey, asking whether each of the presented images was a real US scan or an ``AI Generated'' (\ie UltraGauss-generated) scan.}
    \label{fig:survey-bonus}
\end{figure}

\begin{figure}
    \centering
    \includegraphics[width=0.3\linewidth]{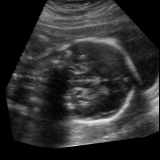} \hfill \includegraphics[width=0.3\linewidth]{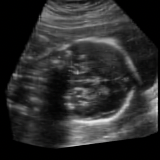} \hfill \includegraphics[width=0.3\linewidth]{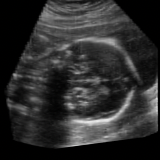}
    \caption{\textbf{Left:} Ground Truth Mid-Coronal Scan, \textbf{Centre:} UltraGauss @ $t=4$ mins, \textbf{Right:} UltraGauss @ $t=15$ mins}
    \label{fig:survey-bonus-imgs}
\end{figure}

\begin{figure}
    \centering
    \includegraphics[width=1\linewidth]{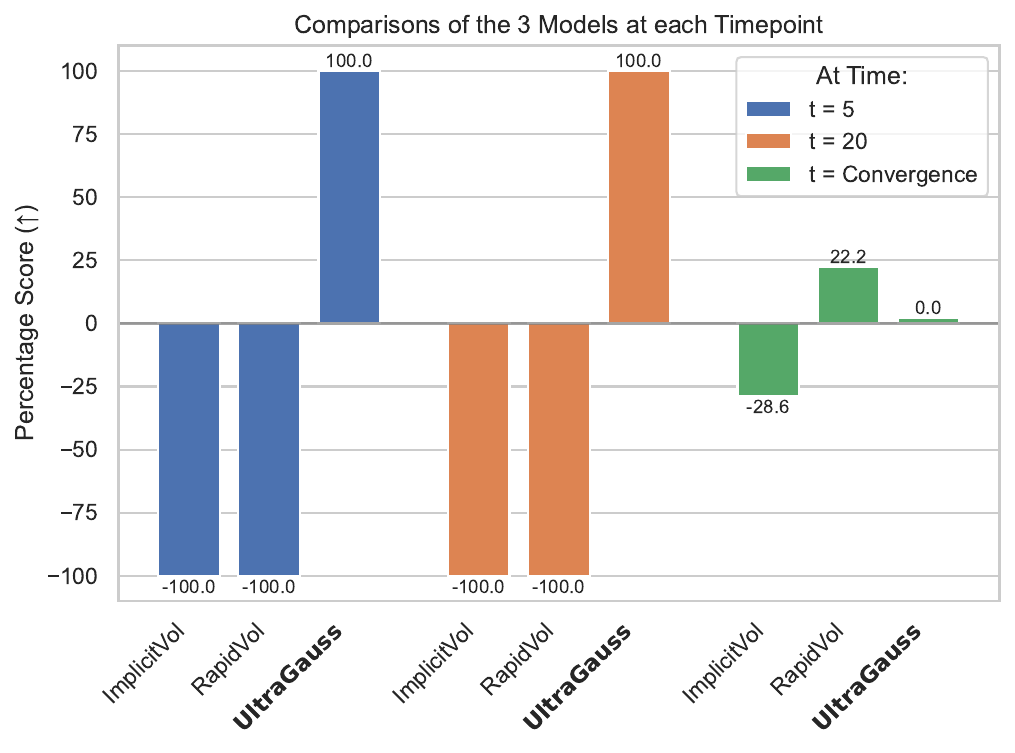}
    \caption{Summary of Survey Results. Actively choosing a model gave it 1 point, not choosing it -1, and choosing ``no preference'' 0 points. Each model's score was then divided by the total number of possible points to be gained, giving a percentage. Thus +100\% means that all participants actively selected and preferred that model, -100\% that all participants actively disliked this model (and selected the other one), and 0\% that the model was neither liked nor disliked, as the participant selected ``no preference''.}
    \label{fig:survey-q1}
\end{figure}

\section{Results}

\begin{figure}
    \centering
    \includegraphics[width=1\linewidth]{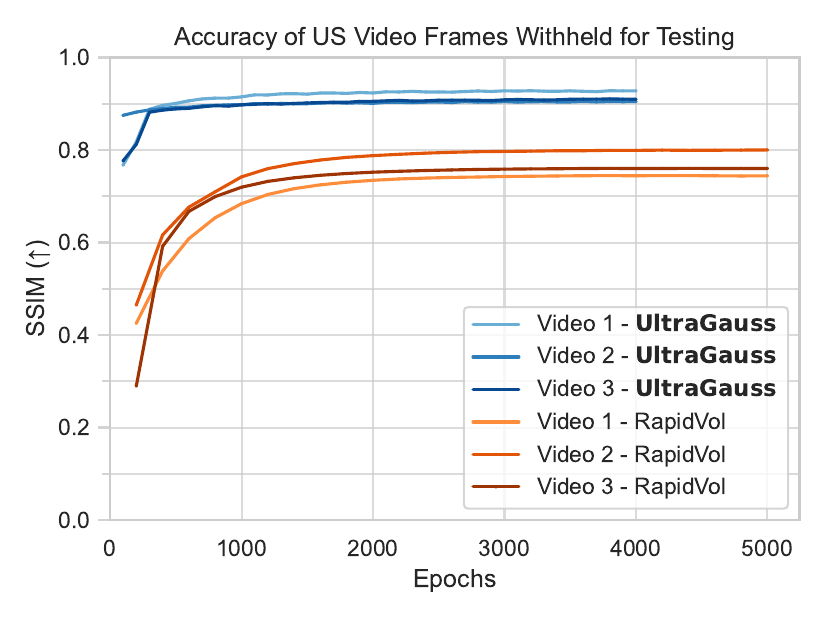}
    \caption{Test accuracy over the course of reconstruction/training for UltraGauss - 300K (ours) and RapidVol, reported on 3 ultrasound cinesweeps. Higher SSIM is better.}
    \label{fig:video exp1 test}
\end{figure}

\begin{table}[t]
\caption{Quantitative results of reconstruction performance on cinesweep videos. The SSIM scores shown are the average across all the frames held-out for testing/used in training at $t=\infty$.\\ Best scores are highlighted in \textbf{bold}. $\uparrow$ indicates higher is better.}
\label{tab:us video results table}
\centering\scriptsize  % small font
\setlength{\tabcolsep}{3pt}  % set column separation
\begin{tabular}{lcccccc}
\toprule
& Model & Video 1 & Video 2 & Video 3 & Avg. & Std. \\
\midrule
\multirow{2}{*}{Test SSIM $\uparrow$}  
& RapidVol          & 0.745  & 0.799  & 0.760  & 0.768  & 0.028 \\
& \textbf{UltraGauss} & \textbf{0.928}  & \textbf{0.905}  & \textbf{0.910}  & \textbf{0.914}  & \textbf{0.012} \\
\midrule
\multirow{2}{*}{Train SSIM $\uparrow$}  
& RapidVol          & 0.871  & 0.884  & 0.880  & 0.878  & \textbf{0.007} \\
& \textbf{UltraGauss} & \textbf{0.959}  & \textbf{0.957}  & \textbf{0.939}  & \textbf{0.952}  & 0.011 \\
\bottomrule
\end{tabular}
\end{table}

\subsection{Quality and Speed Evaluation}
We extensively tested using 3 different test sets, each getting progressively harder and more life-like. The full results table can be found in Appendix A, however the table is nicely summarised in \cref{fig:main experiemnt results graph}. One can clearly see that UltraGauss is more accurate than current SOTA regardless of the test set, and at all 3 timepoints. If near real-time reconstruction/training is desired, then UltraGauss greatly surpasses current state of the art (by at least 0.2 SSIM at $t=5$). It also has near perfect accuracy, with UltraGauss - 2M achieveing an average SSIM of 0.995 upon completion. The cross-age variance is also much smaller (at least $10 \times$ so) than RapidVol and ImplicitVol. One important conclusion is that at each time point, there is an optimum number of gaussians to be had. For example if it is know that recosntruction can only take $t=5$ minutes, then it is preferable to have 100K gaussians, whilst if $t=\mathcal{O}(hours)$ than 2M gaussians would lead to a slightly higher accuracy. However we found that UltraGauss-300K is a good all-round model.

\subsection{Clinicians' survey}
\textbf{Survey participation:} The survey was distributed to 12 expert sonographers specializing in fetal, pediatric, and general ultrasound imaging, from hospitals in four countries (UK, Ghana, Denmark, and the NL). 
Ten experts who regularly interpret ultrasound scans responded, including consultant fetal surgeons and senior sonographers, with an average of 18 years of medical experience (range: 7–30 years).

\textbf{Comparison of reconstruction methods:} UltraGauss consistently outperformed RapidVol and ImplicitVol across all training time points, with \underline{all} clinicians actively preferring UltraGauss over RapidVol or ImplicitVol (see \cref{fig:survey-q1}), especially when quick reconstructions are needed (\ie $t\leq20$). As training progressed, UltraGauss-2M became preferred over UltraGauss-100K and UltraGauss-300K. This indicates that higher-capacity UltraGauss models produce more realistic reconstructions. 

\textbf{Realism of UltraGauss' reconstructions over time:} In the progressive reconstruction experiment, 70\% of the experts rated UltraGauss-generated images as more realistic than the ground-truth after just 4 minutes of training. 
This percentage increased to 80\% after 15 minutes, demonstrating the rapid convergence of UltraGauss towards clinically realistic ultrasound textures (\cref{fig:survey-bonus}). These images can be seen and compared to the ground truth, just as the clinicians had to do, in \cref{fig:survey-bonus-imgs}.
The inclusion of two identical ground-truth images, one of which was not identified as being real by 40\% of the participants, confirmed that even expert clinicians exhibit variability in assessing ultrasound realism, reinforcing the robustness of UltraGauss' ratings. 

\subsection{End-to-end pipeline results}
\cref{tab:us video results table} shows the quantitative results of the images generated at the predicted poses of the video frames withheld for testing, and \cref{fig:video exp1 test} over time. It is evident from both that UltraGauss greatly outperforms SOTA, especially if reconstructions are needed very quickly. Such cases could be in emergency situations, or if consultations have to be kept short. For a task which is as close to real-world usage as it can get, and considering the error is the sum of the pose prediction and reconstruction error, an overall average test SSIM of 0.91 is highly respectable. Appendix C shows this qualitatively. 
\section{Conclusion}
We proposed UltraGauss, a method for 3D reconstruction of ultrasound volumes based on gaussian splatting. We explored the differences in image formation models between light-based imaging and ultrasound imaging, and proposed an intersection-based gaussian rasterization to better suit this setting.
We mathematically derive efficient rasterization boundaries, with an associated parallelization scheme, and a new parametrization of covariances with several appealing properties.
Our experiments demonstrate that gaussian splatting, with the appropriate changes, is an extremely fast and accurate model for ultrasound volumes, with clinical relevance checked by a survey of expert clinicians.
We hope that this work forms the basis for more accessible ultrasound imaging, through 3D reconstruction tools that lower the barrier to interpretation. We also hope that it facilitates future research into diagnosis and prognosis using the gaussian-based 3D modality.
{
    \small
    \bibliographystyle{ieeenat_fullname}
    % \bibliography{main}
    \bibliography{marks_zotero_static, references}
}
\onecolumn %For Supplementary, use one column instead
\appendix

\newpage
\begin{landscape}
\section*{Appendix A: Experiment 5.2 - Results Table}
\begin{figure}[h!]
    \centering
    \includegraphics[width=1\linewidth]{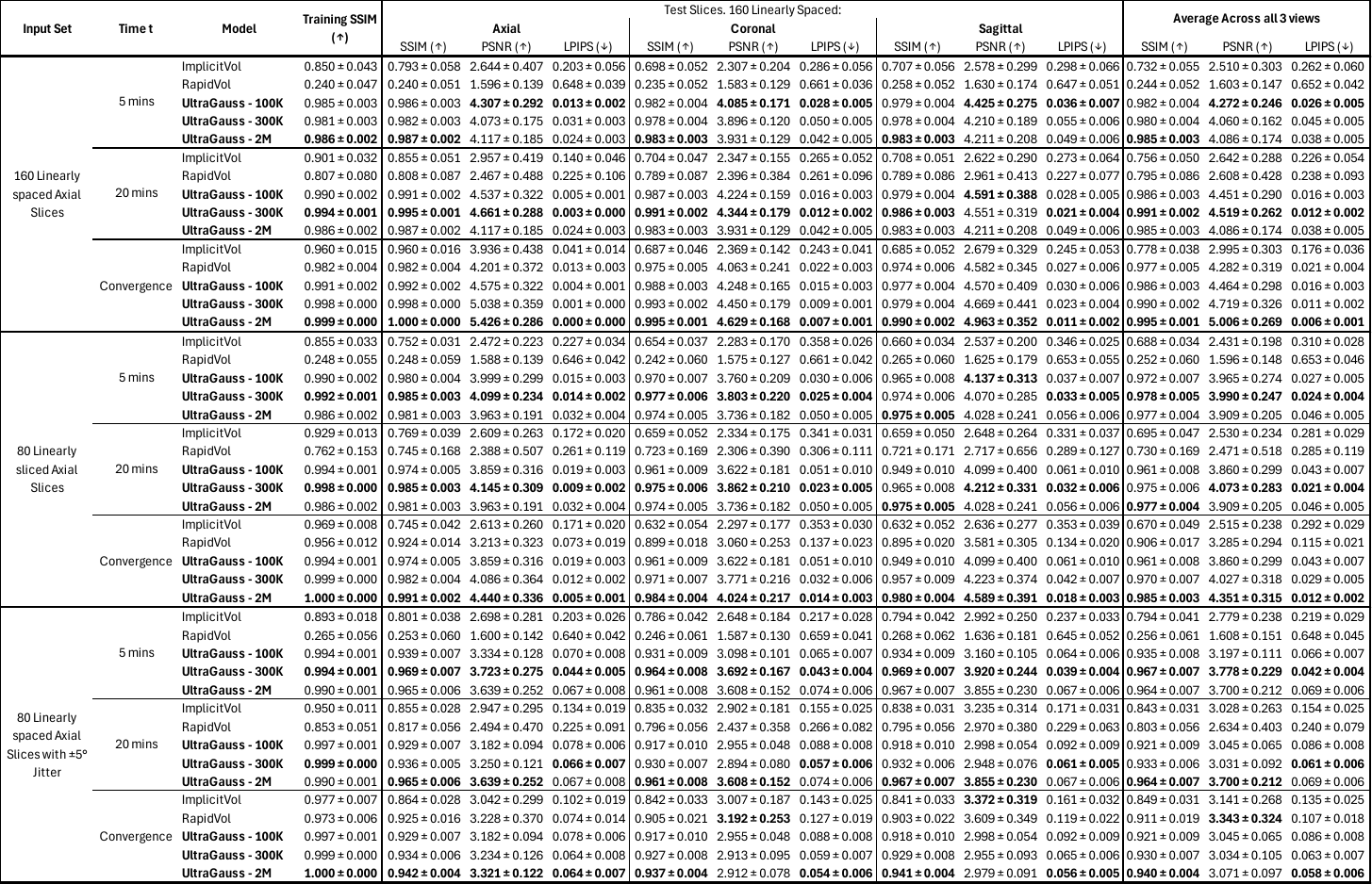}
    \caption{Reconstruction results for 5 different models on 3 different training sets, shown at 3 time points. $\uparrow$ indicates higher is better. ``UltraGauss - $N$'' indicates UltraGauss with $N$ starting gaussians.}
    \label{fig:Appendix_video exp1 test}
\end{figure}
\end{landscape}

\newpage
\section*{Appendix B: Experiment 5.3 - Clinicians' Survey Extra Results}
\begin{figure}[h]
    \centering
    \includegraphics[width=0.7\linewidth]{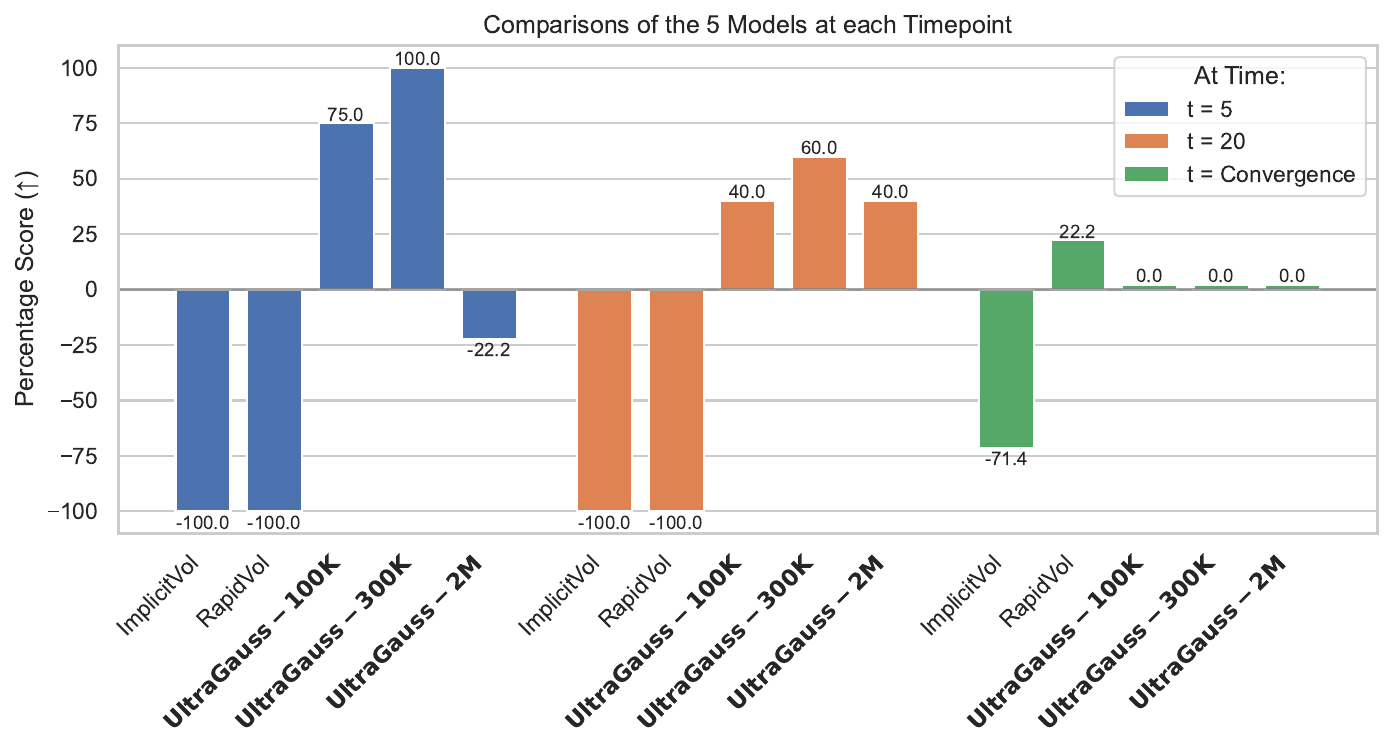}
    \caption{Survey results: Comparison of all 5 models at each time point. Actively choosing a model gave it 1 point, not choosing it -1, and choosing ``no preference'' gave it 0 points. Each model's score was then divided by the total number of points it could have gained, to give a percentage. Thus +100\% means that all participants actively selected and preferred that model. -100\% means that all participants actively disliked this model (and selected the other one). 0\% means that the model is neither liked or disliked, as the participant chose the ``no preference'' option.}
    \label{fig:Appendix_survey-q1}
\end{figure}

\begin{figure}[h]
    \centering
    \includegraphics[width=0.65\linewidth]{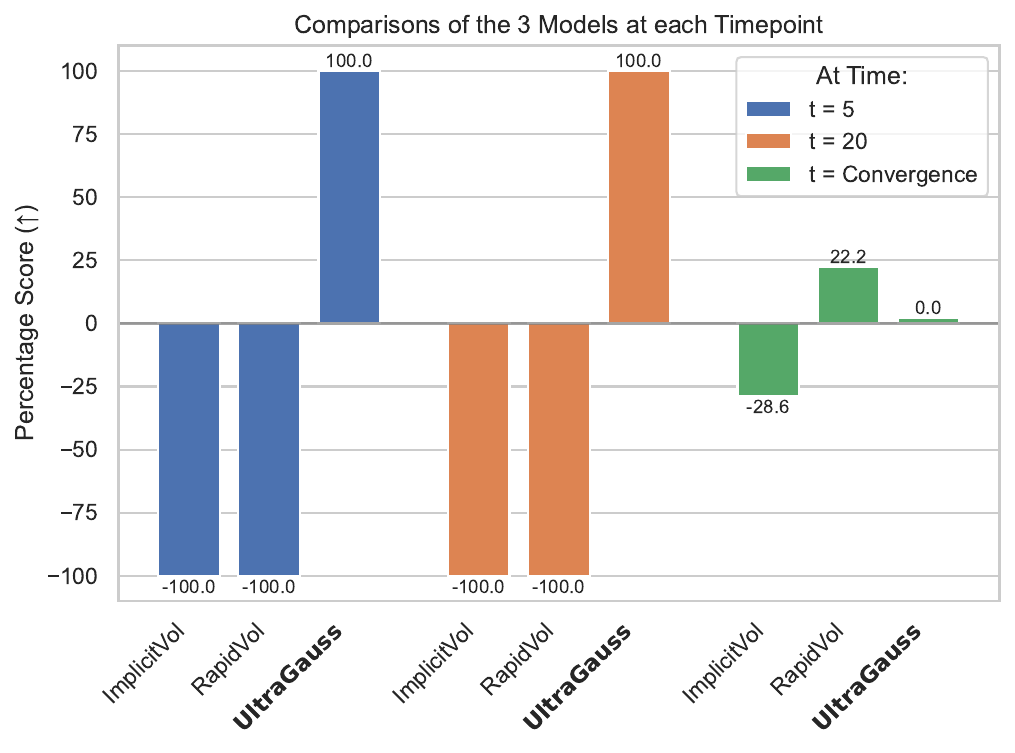}
    \caption{Survey results: Comparison of ImplicitVol, RapidVol and the better of the three UltraGauss Models, at each time point. The same scoring system applies as \cref{fig:Appendix_survey-q1}}
    \label{fig:Appendix_survey-q2}
\end{figure}

\newpage
\section*{Appendix C: Experiment 5.4 - Visual Comparison of Reconstructed Images from Cinesweeps}
\begin{figure}[h] % Use figure* to span both columns
    \centering
    \renewcommand{\thesubfigure}{} % Remove numbering if not needed
    \setlength{\tabcolsep}{2pt} % Adjust spacing between columns
    \renewcommand{\arraystretch}{1} % Adjust row height for better spacing

    \begin{tabular}{c|c|c|c|c|c|c|} % 6 columns including the headers
        \hline
         & \textbf{V1 I1} & \textbf{V1 I2} & \textbf{V2 I1} & \textbf{V2 I2} & \textbf{V3 I1} & \textbf{V3 I2} \\
        \hline
        \raisebox{6ex}{\textbf{Ground Truth}} & 
        \subfloat{\includegraphics[width=0.13\textwidth]{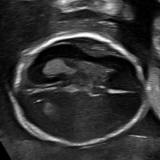}} &
        \subfloat{\includegraphics[width=0.13\textwidth]{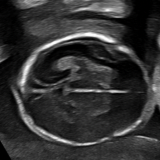}} &
        \subfloat{\includegraphics[width=0.13\textwidth]{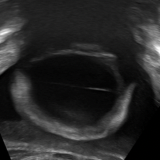}} &
        \subfloat{\includegraphics[width=0.13\textwidth]{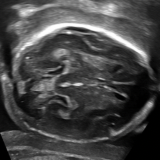}} &
        \subfloat{\includegraphics[width=0.13\textwidth]{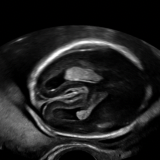}} &
        \subfloat{\hspace{-0.2cm} \includegraphics[width=0.13\textwidth]{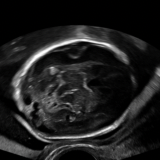}} \\
        \hline
        \raisebox{6ex}{\textbf{RapidVol}} & 
        \subfloat{\includegraphics[width=0.13\textwidth]{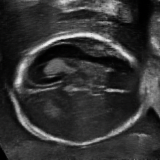}} &
        \subfloat{\includegraphics[width=0.13\textwidth]{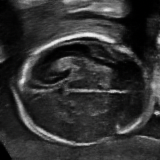}} &
        \subfloat{\includegraphics[width=0.13\textwidth]{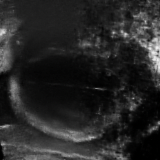}} &
        \subfloat{\includegraphics[width=0.13\textwidth]{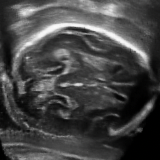}} &
        \subfloat{\includegraphics[width=0.13\textwidth]{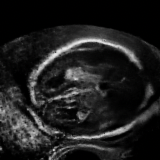}} &
        \subfloat{\hspace{-0.2cm} \includegraphics[width=0.13\textwidth]{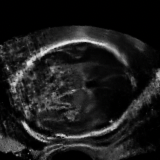}} \\
        \hline
        \raisebox{6ex}{\textbf{UltraGauss}} & 
        \subfloat{\includegraphics[width=0.13\textwidth]{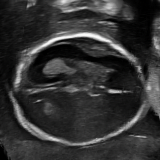}} &
        \subfloat{\includegraphics[width=0.13\textwidth]{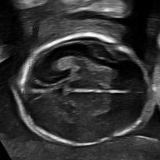}} &
        \subfloat{\includegraphics[width=0.13\textwidth]{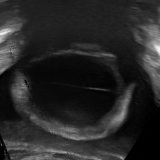}} &
        \subfloat{\includegraphics[width=0.13\textwidth]{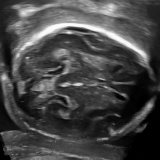}} &
        \subfloat{\includegraphics[width=0.13\textwidth]{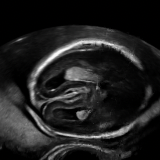}} &
        \subfloat{\hspace{-0.2cm} \includegraphics[width=0.13\textwidth]{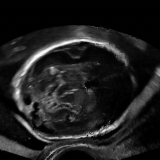}} \\
        \hline
        \raisebox{6ex}{\textbf{RapidVol Diff.}} & 
        \subfloat{\includegraphics[width=0.13\textwidth]{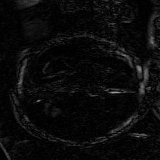}} &
        \subfloat{\includegraphics[width=0.13\textwidth]{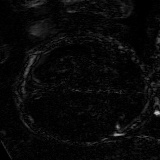}} &
        \subfloat{\includegraphics[width=0.13\textwidth]{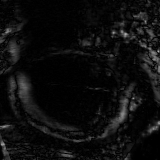}} &
        \subfloat{\includegraphics[width=0.13\textwidth]{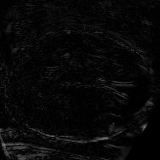}} &
        \subfloat{\includegraphics[width=0.13\textwidth]{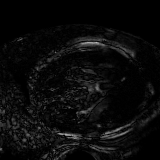}} &
        \subfloat{\hspace{-0.2cm} \includegraphics[width=0.13\textwidth]{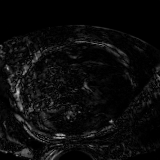}} \\
        \hline
        \raisebox{6ex}{\textbf{UltraGauss Diff.}} & 
        \subfloat{\includegraphics[width=0.13\textwidth]{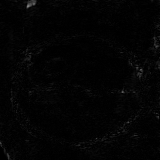}} &
        \subfloat{\includegraphics[width=0.13\textwidth]{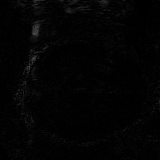}} &
        \subfloat{\includegraphics[width=0.13\textwidth]{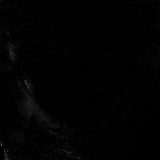}} &
        \subfloat{\includegraphics[width=0.13\textwidth]{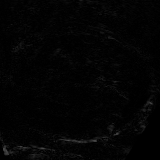}} &
        \subfloat{\includegraphics[width=0.13\textwidth]{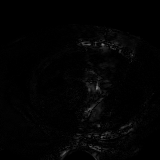}} &
        \subfloat{\hspace{-0.2cm} \includegraphics[width=0.13\textwidth]{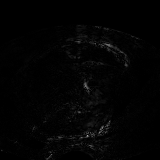}} \\
        \hline
    \end{tabular}

    \caption{For each video (V$n$), two ground truth images (I$j$) from the test set are shown, alongside the corresponding image generated by RapidVol and UltraGauss - 300K (ours). The absolute difference between the predicted and ground truth image is also shown. A pure black diff. image means perfect similarity and so the darker the Diff image, the better.}.
    \label{fig:Appendix_video exp1 images}
\end{figure} %ONLY FOR ARXIV VERSION

\end{document}